\documentclass{ws-ijmpa}
\newcommand{\be}{\begin{equation}}
\newcommand{\ee}{\end{equation}}
\begin{document}

\markboth{R. Narayanan}
{Large N QCD -- Continuum Reduction and Chiral Condensate}

%
\catchline{}{}{}{}{}
%

\title{Large N QCD -- Continuum Reduction and Chiral Condensate}

\author{\footnotesize RAJAMANI NARAYANAN}

\address{Department of Physics, Florida International University, University Park\\
Miami, FL 33199,
USA}

\maketitle


\begin{abstract}
Continuum reduction in large N QCD enables one to extract physical
quantities in the $N\rightarrow\infty$ limit of QCD by working
in small physics volumes. The computation of chiral condensate
is an example of such a calculation.

\keywords{Large N QCD; Continuum Reduction; Chiral Condensate.}
\end{abstract}

\section{Introduction}
QCD with massless quarks
does not have any free parameters. 't Hooft~\cite{hooft1} pointed out
that one might consider the size of the gauge group, N, as a parameter.
The initial hope was to exactly solve QCD in the $N\rightarrow\infty$
limit and compute physical quantities as a power series in $1/N$.
Although the exact solution of large N QCD is not yet available,
significant progress has been in the area of large N phenomenology
by a careful study of large N counting~\cite{lebed}.  
This analysis has extensively shown that $1/N$ is a good
expansion parameter and that many experimental results can be 
reproduced by just studying one or two leading terms in $1/N$.
In addition lattice studies of pure gauge theories~\cite{teper} have shown
that large N limits of quantities like string tension and
deconfinement temperature are reached as early as $N=5$ and $N=6$.
If one also solves the theory in the $N\rightarrow\infty$ limit, then
one can compute physical observables as a power series in $1/N$
with no other free parameter.

It seems quite likely that one can solve for the 
meson spectra in the large N limit of QCD using existing
numerical techniques and moderate computing power~\cite{knn,nn1,knn1,cond}. 
The basic idea behind the numerical solution is the concept
of continuum reduction. Continuum reduction enables one to work
in a finite physical volume and compute observables in the infinite
volume theory without any finite volume effects.
One first takes the $N\rightarrow\infty$ limit at a fixed physical
volume and the resulting theory does not depend on the physical
volume.
The numerical study of large N QCD is further simplified by the
fact that fermions in the fundamental representation are naturally quenched.
This is the case as long as the theory only has a finite number of
fermion flavors in the fundamental representation.

\subsection{Continuum reduction}

Eguchi and Kawai~\cite{ek} made the observation that the space-time lattice
in the limit of $N\rightarrow\infty$ can be reduced to a single
point provided the global $U^d(1)$ symmetries that multiplies
Polyakov loops by $U(1)$ phases are not broken. This symmetry
is broken for $d>2$ and the equivalence of the loop equations
used by Eguchi and Kawai is ruined~\cite{bhn}. But, the arguments of Eguchi and
Kawai also hold for an $L^d$ lattice where $L >1$. That is to say,
the infinite space-time lattice can be reduced to a finite $L^d$
lattice in the limit of $N\rightarrow\infty$ provided the
global $U^d(1)$ symmetries that multiplies
Polyakov loops by $U(1)$ phases are not broken on the $l^d$ lattice.

The only parameter in the lattice theory is the 't Hooft gauge coupling
$b=\frac{1}{g^2_0N}$. 
The location of the transition point $b_c(L)$
where one of the $U^d(1)$ symmetries is broken 
scales properly with $L$~\cite{nn1,knn1}.
One can therefore define a critical size
$l_c$ in the continuum such that the $U^d(1)$ symmetries remain
unbroken for $l>l_c$. The argument of Eguchi and Kawai will hold for
all $L$ as long as we keep $b<b_c(L)$ and there will
be no dependence on $L$. There will be no dependence on the box size $l$
as long as $l > l_c$ and this is referred to as continuum reduction.
Physical
results on the lattice
can be extracted by working on an $L^d$ lattice and keeping
$b$ just below $b_c(L)$. Computations should be done on two or three
different values of $N$ to study the infinite $N$ limit. Once this
is done, it is sufficient to work with two or three different $L$
values to study the effect of finite lattice spacing and one will
be able to extract the results for infinite volume QCD in the limit
of large $N$. 

There is one lattice artifact to be taken into account when setting
the parameters $\{L,b<b_c(L)\}$. There is a bulk transition on the
lattice in the large $N$ limit at $b_B=0.36$
that is associated with the spectrum
of the single plaquette ($1\times 1$ Wilson loop).\footnote {There
is no phase transition for finite $N$ but there is a cross-over
that gets stronger with increasing $N$.} 
One has to set $b> b_B$ to obtain
the proper continuum limit. 
The scaling of the critical coupling $b_c(L)$ in perturbation theory 
up to two loops is
given by
\be
\Lambda L = \Biggl[\frac{11}{48\pi^2b_c(L)}\Biggr]^{\frac{51}{121}} 
e^{\frac{24\pi^2b_c(L)}{11}}\label{twoloop}
\ee
Lattice study of the phase transtion where one goes from $0c$ 
(phase where $b > b_B$ and all $U^d(1)$ are unbroken) to
$1c$ (phase where $b > b_B$ and one of the $U^d(1)$ is broken) shows that
a value of $\Lambda=3.85\pm 0.2$ in Eqn.(\ref{twoloop})
fits the data well with $b_c(L)$ replaced by its tadpole improved
value.

\section{Chiral condensate}
QCD with massless quarks has a chiral symmetry that implies 
$\langle\bar\psi\psi\rangle = 0$ as long as one is in a finite
physical volume. But the massless limit and the infinite volume do
not commute. In particular,
\be
\lim_{m\rightarrow 0}\lim_{V\rightarrow\infty} \frac{1}{V} \langle\bar\psi\psi (m) \rangle 
=\Sigma \ne 0
\ee
in QCD at zero temperature indicating that chiral symmetry is spontaneously broken.
The theory does not depend on the physical volume in the large N limit as long as
$l > l_c$. Therefore, the massless limit has to commute with the infinite volume
limit. How does chiral symmetry break in large N QCD? The answer lies in the
$N\rightarrow\infty$ limit. This limit does not commute with the massless limit.
In particular,
\be
\lim_{m\rightarrow 0}\lim_{N\rightarrow\infty} \frac{1}{Nl^4} \langle\bar\psi\psi (m,N,l) \rangle 
=\Sigma\ne 0
\ee
and this $\Sigma$ does not depend on $l$ as long as $l>l_c$.

Overlap fermions~\cite{over} enable one to study chiral symemtry breaking
away from the continuum limit.
The computation of the
chiral condensate was performed as follows. Gauge fields were generated
at a fixed $L$, $N$ and coupling $b=\frac{1}{g^2N}$. Gauge fields
were generated in the $Q=0$ and $Q=1$ topological sectors. 
Two lowest non-zero eigenvalues of the massless hermitian
overlap Dirac operator, $\lambda_1$
and $\lambda_2$, were computed on each configuration.\footnote{More eigenvalues can
be computed but this increases the computational cost and two are sufficient to
obtain an estimate of the chiral condensate.}
If chiral symmetry is spontaneously broken as $N\rightarrow\infty$
at a fixed $L$ and $b$, then the distributions
of $z_i=\lambda_i \Sigma N L^4$ should be given by universal functions
described by chiral random matrix theory~\cite{rmt}. It was shown that
the distribution $p(z_1/z_2)$ was universal as predicted
by chiral random matrix theory as $N$ gets large at a fixed 
$L$ and $b$. Furthermore, we determined $\Sigma$
such that both $p_1(z_1)$ and $p_2(z_2)$ followed universal 
distribution functions. We showed that $\Sigma$ did not depend on
$L$ after one has taken the large N limit at a fixed $b$. This
confirmed that continuum reduction hold for fermionic operators.
 Furthermore, $\Sigma(b)$ obeyed the usual scaling law of QCD
and $\Sigma(b) L^3_c(b)$ was a constant. The detailed analysis
resulted in the following estimate for the chiral condensate.
\be
\frac{l_c^3}{N_c} \langle \bar\psi\psi \rangle^{\overline{MS}}(2 GeV) \approx (0.65)^3
\ee

\section{'t Hooft model}

It is interesting to consider two dimensional QCD where
one has a chiral condensate. Continuum reduction is trivial in
$d=0$ since $l_c=0$. Therefore, one should be able to extract the chiral
condensate by working on a $1^2$ lattice. The only degrees of freedom are
two SU(N) matrices we will call $U$ and $V$. 
The $2N\times 2N$ hermitian overlap Dirac matrix, $H_o$ 
is a dense matrix arising out of the denseness of $U$ and $V$.
This model looks very close to a chiral random matrix model
with the complication arising out of the fact that one has two $N\times N$ matrices.
We are not aware of an exact solution of this model that results in 
universal distributions of the eigenvalues of $H_o$.
Let $\pm \lambda_i$, $i=1,\cdots, N$ be the $2N$ eigenvalues of $H_o$.
Then the joint disitribution of
$z_i = \frac{\lambda_i N}{\sqrt{6\pi b}}$ should obey the universal functions
given by chiral RMT since $\Sigma =\frac{1}{\sqrt{6\pi b}}$.

\begin{figure}
\centerline{\psfig{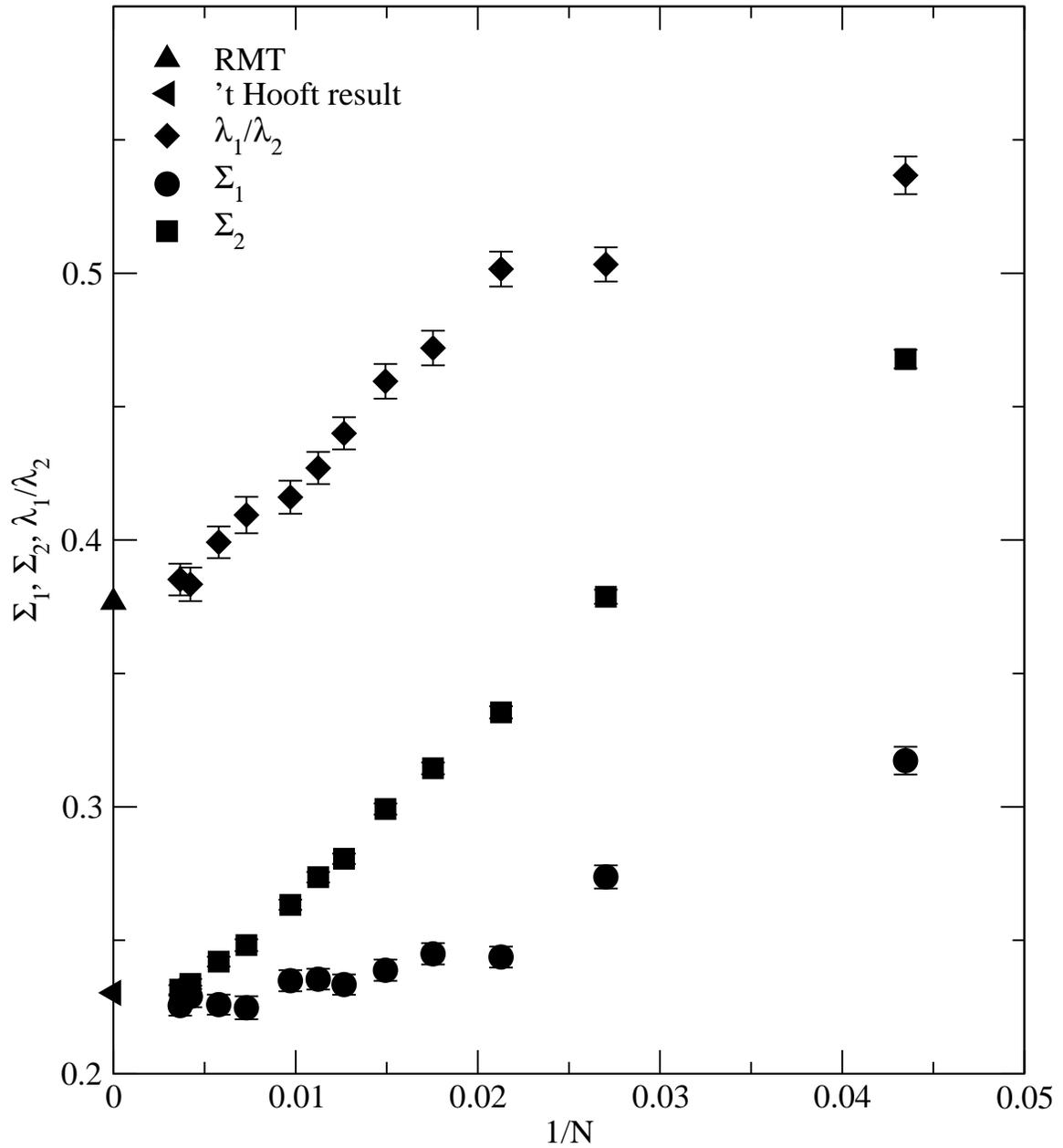}}
\vspace*{8pt}
\caption{Extraction of $\Sigma_1$ and $\Sigma_2$ from numerical simulations
along with the average value of $r=\frac{\lambda_1}{\lambda_2}$ as a function
of $N$.}
\label{fig1}
\end{figure}

\begin{figure}
\centerline{\psfig{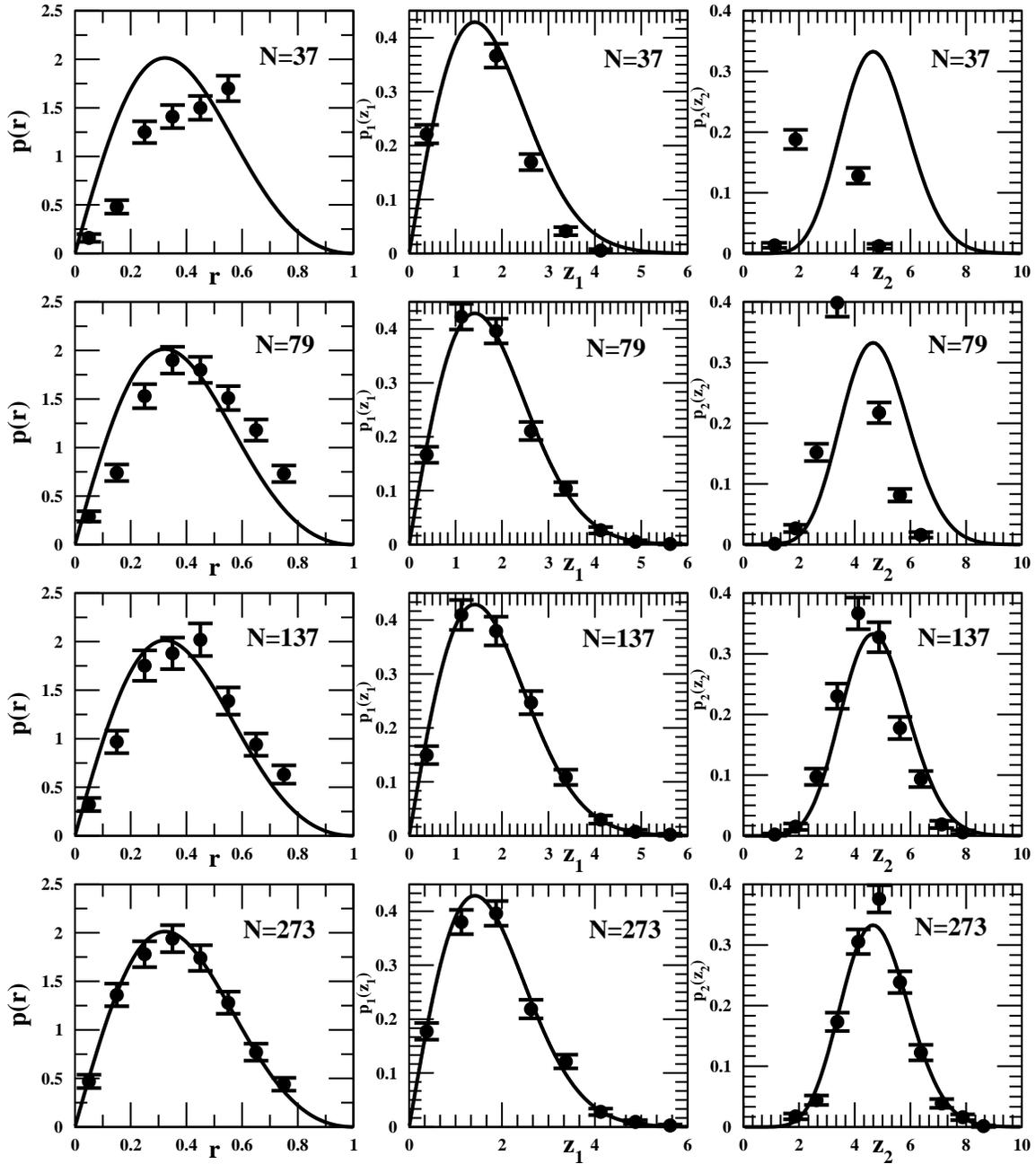}}
\vspace*{8pt}
\caption{Comparison of the distributions of the scaled eigenvalues $z_1$ and
$z_2$ with the chiral RMT distributions for the same. Also shown is
the comparison of $p(r)$ with that of chiral RMT.}
\label{fig2}
\end{figure}

We performed a numerical simulation of this model for $b=1$ and several
different $N$ values. We computed all eigenvalues of $H_o$ and obtained
an estimate of $\Sigma_i$ from the two lowest eigenvalues by forcing the
average of $z_i$ to be the same as the one dictated by chiral RMT. 
Fig.~\ref{fig1} shows $\Sigma_1$ and $\Sigma_2$ as a function of $N$
and one can clearly see the approach to the correcy value as $N\rightarrow\infty$.
Furthermore, the average of the ratio, $r=\frac{\lambda_1}{\lambda_2}$,
also approaches the correct value as shown in Fig.~\ref{fig1}.
Fig.~\ref{fig2} shows the distribution of $p_1(z_1)$, $p_2(z_2)$ and
$p(r)$ obtained from the numerical simulation and the comparison with
the universal distributions given by chiral RMT. The distributions approach
the expected results as $N\rightarrow\infty$.

\section{Conclusions}

Continuum reduction in the large N limit of QCD enables one to
compute physical observales at infinite volume
by performing numerical simulations
on finite physical volumes.  Computation of chiral
condensate was used as an example of such a calculation.
It would be nice to obtain an analytic derivation of
the connection between chiral RMT and 't Hooft model
shown by numerical simulations in section 3.

\section*{Acknowledgments}
Work reported here has been done in collaboration with
Joe Kiskis and Herbert Neuberger.
The author would like to thank Richard Brower and Chung-I Tan for the
invitation to speak at the $8^{\rm{th}}$ workshop on Non-perturbative QCD.
The author acknowledges partial support by the NSF under
grant number PHY-0300065 and also partial support from Jefferson 
Lab.

\end{document}